\documentstyle[12pt,aaspp4]{article}
\begin{document}

\title{Response to the Document ``Origin of the Extended EUV Emission 
from the Abell 2199 and Abell 1795 Clusters of Galaxies''
by Lieu, Mittaz, Bonamente, Durret and Kaastra}

\author{S. Bowyer, T. Bergh\"ofer, and E. Korpela}

Lieu, Mittaz, Bonamente, Durret and Kaastra (hereafter, Lieu et al.) have 
provided a document which claims to rebut the finding by Bowyer, Bergh\"ofer and 
Korpela (hereafter, BBK) presented at the Ringburg Workshop (April 1999) that excess EUV emission detected in some clusters of 
galaxies is in an artifact of the background subtraction employed. 

We invite interested observers to carry out this analysis for themselves, but 
we realize this may take a substantial effort, and not everyone will have the 
necessary tools readily available.  Hence we here provide some relevant 
information and discuss the points raised by Lieu et al. 

The central point is that the EUVE Deep Survey Telescope response is not 
flat. This is not unique to EUVE. All space borne and ground based telescopes 
show this feature to some extent and data from these telescopes are routinely 
corrected with a vignetting function, or flat-fielding. In Fig.~1 below, we 
provide a sensitivity plot of the EUVE Deep Survey Telescope obtained by long 
observations of blank sky. Each contour is a 10\% sensitivity level change. A 
detailed version of this response, useful for considering small-scale variations in the
detector that might be thought of as variations in the detailed cluster emission, is 
available at {\em http://sag-www.ssl.berkeley.edu/$\sim$korpela/euve\_eff}.

\begin{figure}[tb]
\begin{center}
\includegraphics{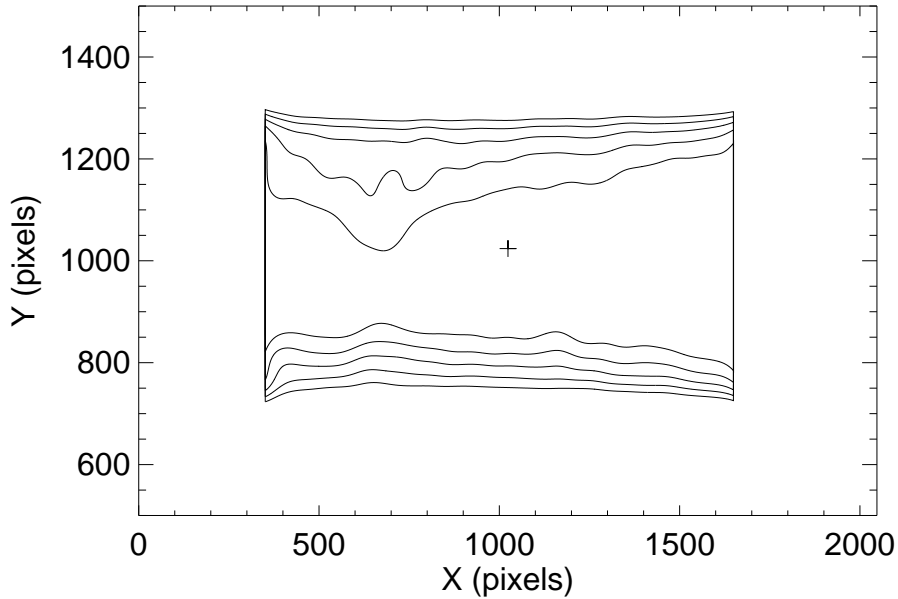}
\vspace{8cm}
\end{center}
\caption{Variations in the sensitivity over the field of the EUVE deep survey
instrument}
\end{figure}

Given the obvious variation in sensitivity over the field of view of this 
telescope, we do not understand how anyone can claim that a flat background as employed by Lieu et al. is 
correct and appropriate for analyses of extended features. 

Nonetheless, in their Figure 3, replicated below, Lieu 
et al. do claim that this background is flat. 
However, it
is visually obvious that the data in this figure are not compatible with a flat background. A 
simple Chi-squared test is something anyone can easily do with the data in this figure; we find a reduced Chi-square of 1.7 for the best fit flat line, and a reduced Chi-square of 0.93 for the best fit sloped line.  
The decrease in the background with increasing radius seen in this figure is 
precisely the effect that we have brought to everyone's attention. 

In their Figure 3, Lieu et al. place their "flat" background level at a rough
average of their data points and claim this to be the average "flat" value.  In their analyses of cluster emission they place their background level near the lowest of their outlying data points.
It is not surprising that by using this method of determining the background
level they find significant positive flux.

\begin{figure}[tb]
\begin{center}
\includegraphics{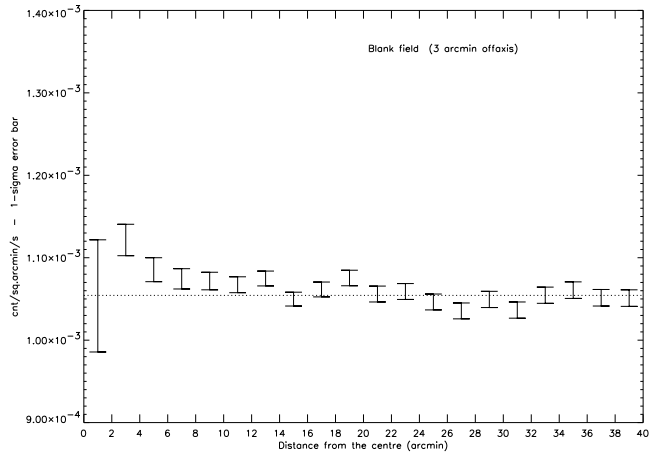}
\vspace{5cm}
\end{center}
\caption{A reproduction of part of Lieu et al. Figure 3}
\end{figure}

Lieu et al. make a number of other points which they claim buttress their 
view. 
\begin{enumerate}
\item They state, "The CSE effect was confirmed by the LECS instrument 
aboard BeppoSax." The validity of this "confirmation" is not clear. The preliminary BeppoSax analysis now available uses theoretical functions for several of the key data reductions, rather than in-flight derived values. A more though analysis is currently underway (Kaastra, private communication).

\item "A multi-scale wavelet analysis of the EUVE/DS data of A 1795 shows 
clear signatures of cluster emission out to a radius of at least 8 arc 
minutes." We find that this "emission" is entirely a fine grained detector 
sensitivity effect. This can be verified by looking at the fine scale 
structure in the background sensitivity map provided at the above listed 
site. The "emission" appears to be different in different clusters because 
the observations were taken at different places on the detector and/or 
were taken with different thresholding and/or were taken at a 
combination of places on the detector. When the detector's small scale 
sensitivity is properly accounted for, this effect disappears.

\item Lieu et al. point out the additional background subtracted by us did not 
lead to the removal of CSE from the Virgo and Coma cluster data. They 
then state, "A natural puzzle is why these two clusters exhibit CSE..." It 
is indeed a puzzle why these two clusters exhibit CSE and others don't. 
The reason for this should be determined by future research. A clue may 
be provided by the fact that these two clusters exhibit substantial activity 
either in the form of merging or the presence of a high energy jet, while the other clusters are quiescent.  

\item Lieu et al. state, "In fact, the rest (of the clusters) suffer from the opposite 
effect; they are strongly intrinsically absorbed." It is hard for us to 
understand why this is a problem.  A similar effect has already been noted in 
X-ray observations of these clusters, and its underlying cause in terms of "cooling flow" gas
was discussed in the presentation by Bowyer at the Ringburg Workshop.

\item The "Clincher test". We have difficulty in understanding all of the 
subtleties provided by Lieu et al. in this section, but their claim that the
two raw data sets are essentially identical is correct.  We stated this at the
Workshop.  Lieu et al. make a number of incorrect statements regarding the
background levels observed.  The particle background cannot be removed by
pulse height thresholding as they claim, it can just be reduced.  They make the  statement that 
different observations can vary by a factor of two "mainly due to an increase
in the photon background" and state that this is a crucial point. Unfortunately this is incorrect. The photon background is constant and the  particle background
is what changes in the EUVE Deep Survey Telescope as was pointed out in an extensive analysis by Lieu and co-workers. ("EUVE First Light Observations of the Diffuse Sky Background",
Lieu, Bowyer, Lampton, Jelinsky, \& Edelstein. 1993, ApJ, 417, L41).

\item It is not true, as Lieu et al. claim, that "within
the context of the BBK scenario, the photon background must assume {\em two}
templates, suitably correlated with each other as to produce the same 
absolute brightness profile."  The template is the same; it is shown in
Figure 1.  The only difference required is a normalization factor
to account for the different (flat) particle background levels at the
time of each observation.

\item The differences in the two data sets shown Figure 6 of Lieu et al. is simply
explained. These observations were taken at different locations on the detector
(as Lieu et al. state).  The vignetting is different at each of these 
locations as can be seen in Figure 1, and hence the profiles will be different.

\item In "Another Cosmic Conspiracy", Lieu et al. state "one will be forced to 
conclude that such a profile must apply to every cluster observed by 
EUV, i.e., all clusters must appear in the EUV like A 2199." We disagree 
with this statement on several grounds. First, if the data were taken at 
different places on the detector, a cursory examination of
Fig.~1 shows that the sensitivity 
deviations will be different and the cluster will look different. Second, 
there is EUV emission from the low energy continuation of the X-ray 
gas in clusters. Since the X-ray gas distribution is different in different clusters, the 
related EUV emission will be different in different clusters. 

\item Lieu et al. incorrectly state that we find no emission in A2199 at radii larger than five
arcminutes.  Our analysis shows that the emission in A2199 extends to at least
9 arcminutes, but is entirely accounted for by the EUV tail of the X-ray emitting gas.

\end{enumerate}

We challenge Lieu et al. to do the following: At each individual position 
where an observation of a cluster is made, derive an azimuthally averaged 
radial profile. Derive an azimuthally averaged radial profile of the background 
taken at this same location.  Subtract from each a particle background as determined by count rates in highly vignetted regions near the edge
of the filter.  Fit the background profile at large radii to the source observations at
large radii, and plot them on the same graph.  Then share the results with
all of us.

\end{document}